\documentclass[conference]{IEEEtran}

\ifCLASSINFOpdf
\usepackage[pdftex]{graphicx}
\else
\fi

\usepackage{cite}
\usepackage{graphicx}
\usepackage{float}
\usepackage{amsmath}
\usepackage[utf8]{inputenc}
\usepackage[final]{pdfpages}
\usepackage{pdfpages}
\usepackage{lipsum}
\usepackage{textcase}
\usepackage{url}
\usepackage{amsmath,esint}
\usepackage{epstopdf}
\usepackage{array}

	\newcommand\blfootnote[1]{%
		\begingroup
		\renewcommand\thefootnote{}\footnote{#1}%
		\addtocounter{footnote}{-1}%
		\endgroup
	}
\newcolumntype{P}[1]{>{\centering\arraybackslash}p{#1}}
\newcolumntype{M}[1]{>{\centering\arraybackslash}m{#1}}

\pdfoutput=1

\begin{document}

\title{Ultra-wideband Channel Modeling for Hurricanes}

\author{\IEEEauthorblockN{Wahab Khawaja\IEEEauthorrefmark{1},
Ismail Guvenc\IEEEauthorrefmark{1}, and
Arindam Chowdhury\IEEEauthorrefmark{2}}
\IEEEauthorblockA{\IEEEauthorrefmark{2}Department of Electrical and Computer Engineering, Florida International University, Miami, FL}
\IEEEauthorblockA{\IEEEauthorrefmark{1}Department of Electrical and Computer Engineering, North Carolina State University, Raleigh, NC}
Email: \{wkhawaj, iguvenc\}@ncsu.edu, \{chowdhur\}@fiu.edu
	
}

\maketitle
\blfootnote{This work has been supported in part by the NSF grants AST-1443999 and CNS-1453678, and W. Khawaja has been supported via a Fulbright scholarship.}
\begin{abstract}
Maintaining communications during major hurricanes is critically important for public safety operations by first responders. This requires accurate knowledge of the propagation channel during hurricane conditions. In this work, we have carried out ultra-wideband (UWB) channel measurements during hurricane conditions ranging from Category-1 to Category-4, generated at the Wall of Wind (WoW) facility of Florida International University (FIU). Time Domain P410 radios are used for channel measurements. From the empirical data analysis in time domain, we developed a UWB statistical broadband channel model for hurricanes. In particular, we characterize the effects of rain and wind speed on large scale and small scale UWB propagation parameters. %To the best of our knowledge, there is no such experiment available in the literature. 

\begin{IEEEkeywords}
Channel measurement, channel modeling, hurricane, Ultrawideband (UWB).
\end{IEEEkeywords}

\end{abstract}

\IEEEpeerreviewmaketitle

\section{Introduction}

Hurricanes are one of most destructive forces in nature where wind speeds exceeding 74 mph and heavy rainfall are observed. An estimated 18 hurricanes of different categories hit United States every decade on the average \cite{NOAA}. Hurricanes can lead to difficulties in existing communication links as the design of general radio transceivers do not take into account such severe conditions. This can make the rescue and relief operations during those conditions extremely challenging. Other sensitive operations such as air traffic control and military communications are also affected in such conditions. With the recent use of autonomous modes of transportation in air, sea and ground, a reliable communication link in all weather conditions is paramount.

Ultra-wideband (UWB) radio signals with their large bandwidth, high data rates, and minimum interference to existing communication links are used in numerous applications. These applications include radars for search, rescue and imaging in the disaster hit areas. A reliable propagation channel model in these conditions such as hurricanes can help in efficient rescue operations. There are UWB channel models available in the literature for indoor and outdoor environments \cite{intel_uwb, molisch00}. However, to our best knowledge, there is no study available to characterize UWB channels in high wind and rain conditions. In \cite{uwb4,uwb6,uwb7,Weather_new1,Weather_new3,Weather_new4} weather effects on communication channel at different frequencies are discussed in controlled and outdoor environments with different foliage concentrations. The wind speeds and rain intensities considered in these studies are much lower than those encountered in a hurricane. 
   
\begin{figure}[!t]
	\centering
	\includegraphics[width=\columnwidth]{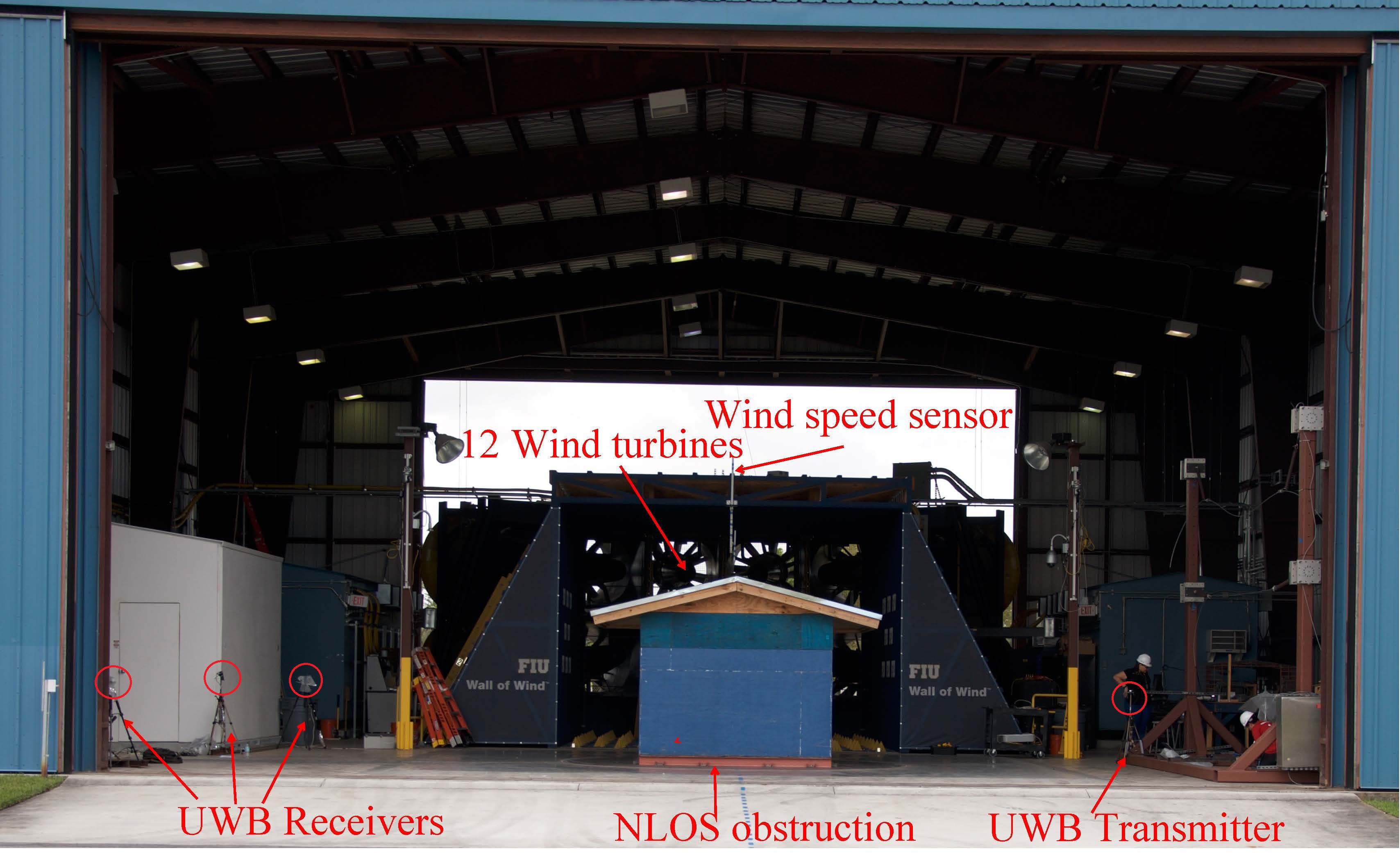}
	\caption{UWB channel measurement setup at Florida International University, Wall of Wind facility, Miami, FL.}\label{Fig:fig_actual}
\vspace{-0.4 cm}
\end{figure}

In this work, we have carried out channel sounding of UWB radio signals in hurricanes generated from Category-1 to Category-4 based on Saffir-Simpson hurricane scale~\cite{SaffirSimpson} at Wall of Wind (WoW) facility in Florida International University (FIU) with the set up shown in Fig.~\ref{Fig:fig_actual}. Time domain P410 radios are used for pulse based channel sounding. The frequency range of operation is 3.1~GHz - 5.3~GHz. The empirical analysis of data in time domain is carried out in order to build a statistical channel model for line-of-sight (LOS), and none-line-of-sight (NLOS) propagation paths in different scenarios during a hurricane. The statistical channel model is found to closely fit the empirical data.

\section{UWB Channel Sounding at FIU WoW}
In this section we summarize the experimental set up at FIU WoW. Special measures were taken in order to protect the communication and data storage equipment from hurricane effects and relaying radio controls from a safe point.  

\subsection{Channel Measurements with UWB P410 Radios}
Time Domain P410 radios in bi-static mode are used in channel measurements due to their ease of setting up and efficiently measuring the channel response in a concise space. The operating frequency range for the experiment is 3.1~GHz - 5.3~GHz. The transmitted power from the radio was limited to -14.5~dBm due to FCC requirements \cite{FCC1}. The radios are configured to send pulses at a rate of 10.1~MHz and for a scan duration of 100~ns. The width of each pulse is 1~ns. A pseudo-random (PR) coded pulse train sequence is used to provide synchronization between the transmitter and receivers eliminating the requirement for a physical connection between the transmitter and the receiver. The PR encoded pulses in the acquisition preamble of the transmitted packet is used to detect and lock the transmitted data at a given receiver. The synchronization clock information is sent from the transmitter to receivers through transmitted packets. The  rake receiver~\cite{guvenc2011reliable} is used to collect the energy in the transmitted waveform at a sampling resolution of 61 ps at the receiving radios.  

\begin{figure}[!t]
	\centering
	\includegraphics[width=\columnwidth]{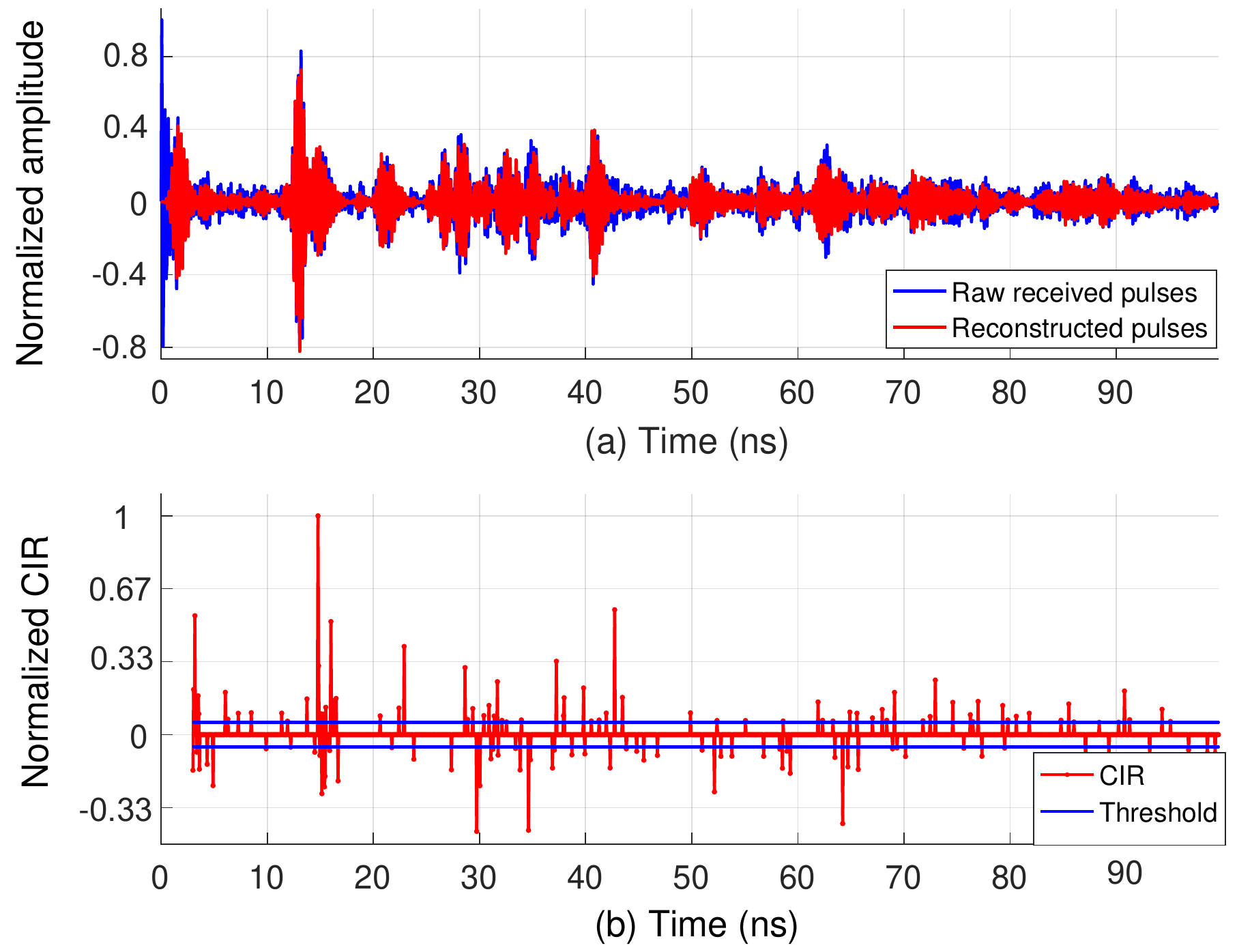}
	\caption{(a) Normalized amplitude of pulses with respect to time for a given scan (blue represents received pulses and red represents reconstructed pulses), (b) Normalized CIR with respect to time for a given scan.}\label{Fig:pulse_tx_rx}
\vspace{-0.4 cm}
\end{figure}

UWB BroadSpec planar elliptical dipole antennas are used in the experiment. The antenna pattern is omni-directional in azimuth direction and doughnut shaped in the vertical direction with gain of 3 dBi. Antennas are placed such that bore-sight of transmitter and receiver antennas face each other giving higher gains \cite{TD}.  The raw received and reconstructed pulses are shown in Fig.~\ref{Fig:pulse_tx_rx}(a). In order to obtain reconstructed pulses, the template waveform is convolved with the channel impulse response (CIR) given in Fig.~\ref{Fig:pulse_tx_rx}(b). The CIR is obtained by deconvolving the received pulses with the template waveform using CLEAN algorithm. The horizontal blue line in Fig.~\ref{Fig:pulse_tx_rx}(b) is the amplitude threshold set at 10\% of the input signal. We have used raw received pulse data in our analysis. 
%\vspace{-0.05 cm}
\subsection{Hurricane Generation Setup}
At the FIU WoW, 12 powerful fans are used to generate hurricanes from Category-1 to Category-5~\cite{SaffirSimpson}. We limited our experiments up to Category-4 hurricane due to limitations of the communication and relaying equipment. Rain is produced through water nozzles placed alongside the fans. The rain intensity in our experiment was set at 223.5 mm/h. 

\begin{figure}[!t]
	\centering
	\includegraphics[width=\columnwidth]{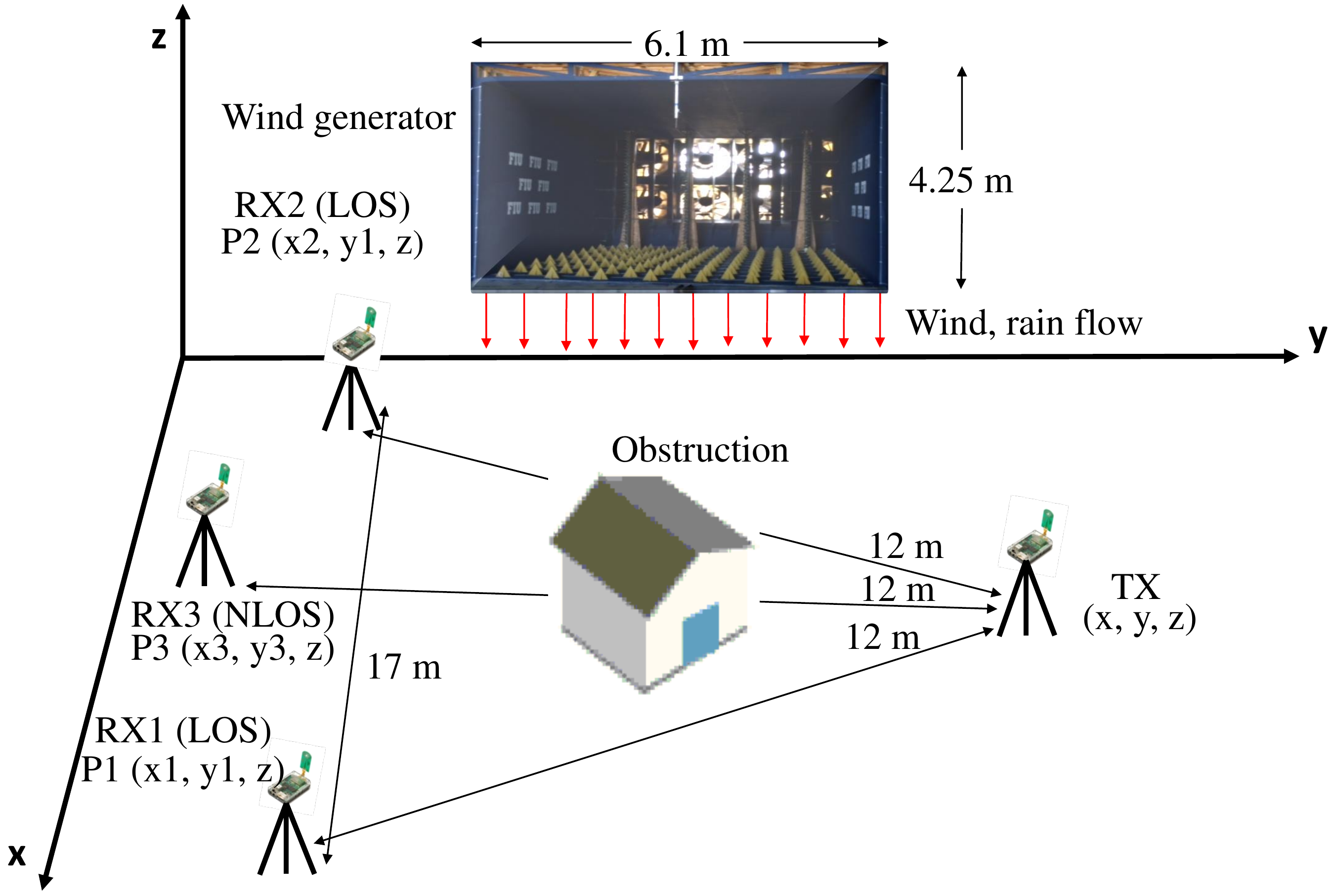}
	\caption{Layout of UWB channel sounding scenario at FIU WoW.}\label{Fig:Exp_Layout}
\end{figure}

The wind pressure due to velocity and density of wind flow at a given spatial location is called pressure head. The density of wind will be considered to be constant along our observation length. From Fig.~\ref{Fig:Exp_Layout}, the position P2 will be under lower pressure head as compared to P1 according to Bernoulli equation along a streamline~\cite{Bernoullieq}. This means that radios at P2 will be exposed to higher wind gusts as compared to at P1. 

In case of wind driven rain (WDR), due to non-uniform motion of water droplets, a plausible assumption is the formation of large sized droplets due to smearing in to each other at higher wind velocities \cite{drop_velocity}. These large sized water droplets have higher likelihood to interfere with the electromagnetic (EM) waves. Due to large frequency band of the EM wave in the experiment, different frequency components will experience different attenuation due to scattering, refraction and diffraction during rain.

Three receivers are placed at equal distance from the transmitter at 12 m as shown in Fig.~\ref{Fig:Exp_Layout}. The LOS radios are at position P1 and P2, and a NLOS path is created by placing a wooden building structure of height 2.76 m between the transmitter and the receiver at position P3. The height of the transmitter and receivers is kept same at 1.5 m. Two scenarios are studied for each receiver radio position labeled as S1 and S2. In S1, we have no rain, whereas in S2, we have rain. In both scenarios, the wind velocity is varied from 90 mph to 140 mph in six discrete steps. To reduce the impact of static objects in the environment, we subtracted the mean statistical CIR at each radio positions without hurricane conditions from the measured ones in hurricane.
 
\section{UWB Channel Modeling for Hurricanes}
In this section, a statistical channel model for UWB radio signals in the frequency band 3.1 GHz - 5.3 GHz is developed for LOS and NLOS paths in hurricane based on the empirical data. We use modified Saleh Valenzuela (SV) channel model \cite{Saleh} for the representation of our UWB channel. The UWB channel in our analysis is considered to be linear time invariant with frequency selective fading.    
\begin{figure}[!t]
  \includegraphics[width=\columnwidth]{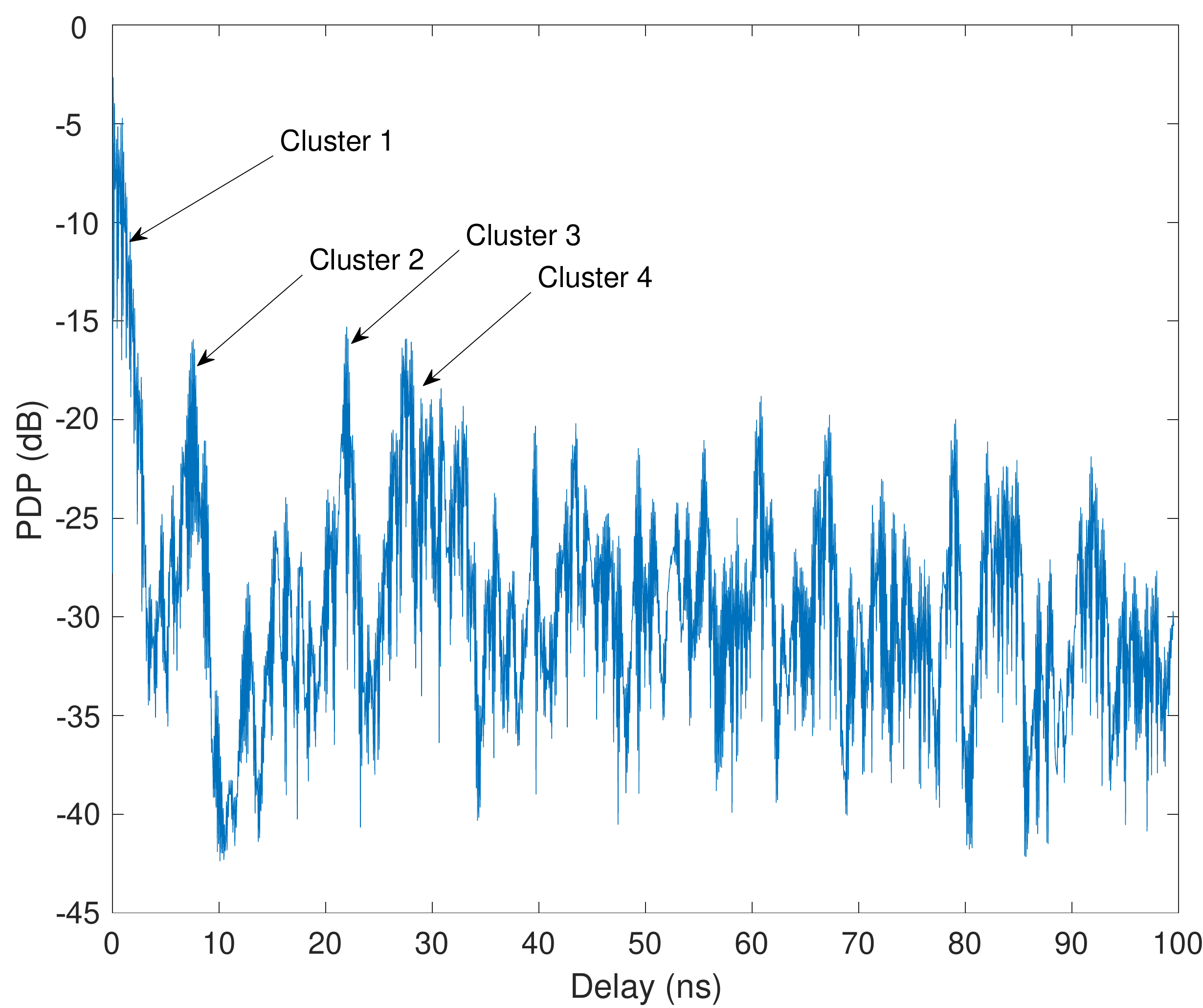}
 \caption{Normalized PDP with respect to delay for a given scan.}\label{Fig:PDP_SV}
\vspace{-0.4 cm}
\end{figure}
\subsection{Multipath Components Analysis}
UWB channel can be characterized based on its multipath components (MPCs) obtained from its CIR:  
\begin{equation}
H(n) = \sum_{i=1}^{N} \sum_{ l=1}^{L}a_{i,l}\exp(j\phi_{i,l})\delta(n-\Gamma_{ i}-\tau_{i,l}) , \label{Eq:CIR_SV}
\end{equation}
where $H(n)$ is the discrete UWB channel response in time domain, $L$ is the total number of MPCs in the $i^{\rm th}$ cluster, $N$ represents the total number of clusters during the scan interval, $a_{ i,l}$, $\phi_{i,l}$, $\tau_{i,l}$ represent the amplitude, phase and delay of the $l^{\rm th}$ MPC in the $i^{\rm th}$ cluster, respectively, and $\Gamma_{i}$ is the delay of the $i^{\rm th}$ cluster. The phase is a uniformly distributed random variable in the interval $[0 \ 2\pi]$, thus it can be neglected. Power delay profile (PDP) obtained using CIR in the LOS case is shown in Fig.~\ref{Fig:PDP_SV}. The formation of major clusters is due to reflection of the MPCs from objects of size comparable or larger than 13 cm. We identify clusters in the PDP using our cluster identification algorithm that is based on the covariance likelihood of the samples of the PDP. The gain amplitudes of MPCs from measurements better fit to the lognormal distribution instead of Rayleigh distribution in the SV model. The amplitudes of channel coefficients can be represented as $ a_{i,l} = e^{Z}  \sim\mathcal{LN}(\mu_{\rm a_{i,l}},\sigma_{\rm a}^2) $ where $Z$ is a normally distributed random variable with mean $\mu_{\rm a_{i,l}}$ and variance $\sigma_{\rm a}^2$, respectively.  

%Envelope of each scan comprising of MPCs follow approximately an exponential distribution. Within each, we have different clusters that follow exponential decay as shown in 

The TOA of clusters and MPCs within each cluster follows two separate Poisson processes. For a given radio position if $\gamma$ and $\zeta$ represent the arrival rate of cluster and MPCs within each cluster, then we have \cite{molisch00,Saleh}:
\begin{align}
p(\Gamma_i|\Gamma_{i-1},\gamma) &= \gamma \exp(-\gamma(\Gamma_{i}-\Gamma_{i-1}))~,\\
p(\tau_{i,l}|\tau_{i,l-1},\zeta) &= \zeta \exp(-\zeta(\tau_{i,l}-\tau_{i,l-1}))~.
\end{align}

If $\overline\Gamma$ and $\overline\tau$ represent the mean arrival time of clusters and MPCs within each cluster, respectively, then the effect of wind velocity and WDR can be given as
\begin{equation}   
\overline\Gamma = (1 + c_{\rm c})\overline\Gamma_{\rm b},  \ \ \ \  \overline\tau  = (1 + c_{\rm m})\overline\tau_{\rm b} \ ,  \label{Eq:TOA_rain_wind}
\end{equation}
where $0\leq c_{\rm c}<1$ and $0\leq c_{\rm m}<1$ are arbitrary constants that are dependent on the rain and pressure head, $c_{\rm c}>c_{\rm m}$, while $\overline\Gamma_{\rm b}$ and $\overline\tau_{\rm b}$ are the mean arrival time of clusters and MPCs within respective clusters for the base case without hurricane. The variation in the mean arrival time of cluster and MPCs within them is represented by two random variables $X_{\rm c}\sim\mathcal{N}(0,\sigma_{\rm c}^2)$ and $X_{\rm m}\sim\mathcal{N}(0,\sigma_{\rm m}^2)$, respectively. 

 If the mean number of clusters at a given receiver radio position is represented by $\overline N$, then we have $\overline N = (1+\frac{c_{j}c_{\rm p}}{c_{r}}) \overline N_{\rm b} $ for $j=1,2,3$, $c_{j}\geq0$ is a constant for given receiver radio position in hurricane, $c_{\rm p}>0$ and $c_{\rm r}>0$ are constants proportional to pressure head and rain, respectively, and $\overline N_{\rm b}$ is the mean number of clusters for the base case without hurricane. If $X_{\rm \overline N}$ is the random variable representing the variation in the mean number of clusters for a given radio position, then we have $X_{\rm \overline N}\sim\mathcal{N}(0,\sigma_{\rm \overline N}^2)$. The formation of clusters is dependent on the channel conditions e.g. during rain we have reduced reflectivity from objects and also EM waves experience higher dielectric constant and loss tangent of water as compared to air resulting in higher attenuation of MPCs and formation of fewer number of clusters.  

%\vspace{0.1 cm}
%\textbf{Remark 3:} From empirical results, the clusters and ray arrival rates decrease with wind velocity, lower pressure head and WDR. The TOA of clusters and their variance has higher value at higher wind velocities, lower pressure head and with WDR.   
%\vspace{0.1 cm}

\subsection{Power Delay Profile}
The PDP can help to characterize power distribution in a given channel as a function of propagation delay. We will consider the general case of non-overlapping clusters as basis for our analysis, where the PDP can be written as \cite{molisch00,Saleh}.
\vspace{-0.1 cm}
\begin{align}
\vspace{-0.8 cm}
P(n) &= \sum_{i=1}^{N}P_{\rm mp}^{i}(n)\exp\big(-\frac{\Gamma_i}{\Lambda}\big)\delta(n-\Gamma_i), \label{Eq:PDP_cluster1}\\
P_{\rm mp}^{i}(n) &= \sum_{l=1}^{L} E(a_{i,1}^2)\exp\big(-\frac{\tau_{i,l}}{\lambda}\big)\delta(n-\tau_{i,l}) \label{Eq:PDP_cluster2} , 
\end{align}
where  $P(n)$ denote the PDP for $N$ clusters, $P_{\rm mp}^{i}(n) $ is the PDP for MPCs of $i$th cluster, while $\Lambda$~and $\lambda$ are the power decay constants for inter-cluster and intra-cluster, respectively, and $ E(a_{i,1}^2) $ is the average power corresponding to first MPC of the $i_{\rm th}$ cluster. The decay constant for inter-cluster case is larger than the intra-cluster case, i.e., $\Lambda>\lambda$. We can represent the variation in $P(n)$  and $P_{\rm mp}(n)$ from (\ref{Eq:PDP_cluster1})(\ref{Eq:PDP_cluster2}) by two zero mean normally distributed random variables $X_{\rm P}\sim\mathcal{N}(0,\sigma_{\rm P}^2)$ and $X_{\rm mp}\sim\mathcal{N}(0,\sigma_{\rm mp}^2)$, respectively.

\subsubsection{WDR effect on PDP}
The PDP is effected by WDR due to phenomenon discussed in Section II-B. If $P^{\rm nr}(n)$ represents the PDP with no rain and $P^{\rm rn}(n)$ represent the PDP in the presence of WDR then we have $P^{(\rm rn)}(n) = \beta P^{(\rm nr)}(n) + X_{\rm R},$
where $0<\beta<1$  is an attenuation constant due to WDR. $X_{\rm R}\sim\mathcal{N}(0,\sigma_{\rm R}^2)$ is a random variable introduced due to WDR. %Due to constant intensity of rain throughout the experiment, variations in $R$  due to different rain intensities is not found. 
%We can derive similar expression for overlapping clusters.

\subsubsection{PDP for NLOS Path}
The effect of wind velocity and WDR are more dominant on the NLOS path as compared to LOS due to long propagation path, absence of dominant power component, and more scattered energy distribution. Let $P^{\rm LOS}(n)$  represents the PDP for LOS and $P^{\rm NLOS}(n)$ is the PDP for NLOS; then we have
\begin{align}
P^{\rm LOS}(n) &= 
\begin{cases}
 B_0 \ \ \ \ \ \ \ \ \ \quad \text{if } \forall \ \Gamma_{i},  \  i=1\\
P^{\rm NLOS}(n) \quad \text{if } \forall \ \Gamma_{i},  \  i\neq 1
\end{cases},\label{Eq:PDP_LOS}\\
P^{\rm NLOS}(n) &= P^{\rm LOS}(n) - B_0 ,\label{Eq:PDP_NLOS}
\end{align}
where $B_0$ represents the cluster energy due to LOS component. This component is absent in the NLOS case, where the energy is distributed into multiple smaller energy clusters without a dominant energy cluster.

%\subsubsection{Auxiliary Parameters}
%From PDP, we can get auxiliary parameters such as mean excess delay (MED) and root mean square delay spread (RMS-DS) and approximate coherence bandwidth 
%\begin{align}
%D_{\rm mean} & = \frac{\sum_{\forall n}^{} n N_{\rm s} P_{\rm noc}(n)}{\sum_{\forall n}^{}P_{\rm noc}(n)}~, \label{Eq:Mean_delay}\\
%D_{\rm rms} &= \sqrt{D_{\rm sq}-D_{\rm mean}^2}~,\label{Eq:RMS_delay}\\
%D_{\rm sq} &= \frac{\sum_{\forall n}^{} (n N_{\rm s})^2 P_{\rm noc}(n)}{\sum_{\forall n}^{}P_{\rm noc}(n)}~,  \label{Eq:m_rms_cb}
%\end{align}
%where $D_{\rm mean}$ denotes the MED, $D_{\rm rms}$ denotes the RMS-DS, $n$ represents the discrete time instance, and $N_{\rm s}$ represents the sampling time at the receiver. Approximate coherence bandwidth (CB) can be found by using the relation $\frac{1}{5D_{\rm rms}}$.

\subsection{Large and Small Scale Fading Parameters}
We consider that the transmitter and the receivers are static, and only wind velocity and WDR are the main sources of variation in the channel. Large scale fading is considered as a measure of attenuation in the received energy during hurricane conditions as compared to a reference energy. From the empirical data, the empirical attenuation due to wind and WDR can be written as    
\begin{align}
A_{\rm e}(v) &= 10\log_{10} \frac{\sum_{\forall n}P^{v_{0}}[n]}{\sum_{\forall n}P^{v}[n]},  \label{Eq:Largescale_1} 
\end{align}
where $P^{v}[n]$ is the PDP at distance $d$ and wind velocity $v$ for a discrete sampling instance $n$ of a scan. On the other hand, $P^{v_0}[n]$ is the PDP at sampling instance $n$ of a scan at $d$ = 12 m, wind velocity $v_0$ of 1.86 mph, ambient temperature of  $25 ^{\circ}$C and standard air pressure of 10.135 N/${\rm cm}^2$. The $ \sum_{\forall n}P^{v}[n]$ represents total energy for different scans in each scenario for wind velocity $v$, while $\sum_{\forall n}P^{v_0}[n]$ represents total energy for different scans of the reference scenario. %PDP is obtained from the CIR of a given scan $k$ as follows, $P^{v,\textit{k}}[n] = |H[n,k]|^2,$ where $~n = 1,...,N_{\rm n},  k= 1,...,N_{\rm tot}$, $H[n,k]$ is the CIR at time instance $n$ of the $k^{\rm th}$ scan, $N_{\rm n} = 100$ ns is the scan duration and $N_{\rm tot}$ is the total number of scans for any given scenario which in our case is 175.

 %From the empirical analysis as shown in Fig.~\ref{Fig:atten_los_nlos},
 We use linear regression to obtain large scale fading parameters for LOS, NLOS paths as follows
\begin{align}
A_{\rm w}(v) & = A_{\rm w_0} + \alpha v + X_{\rm A},  \label{Eq:Largescale} 
\end{align}
where $A_{\rm w}(v)$ is the attenuation as a function of wind velocity in the presence and absence of rain at a given distance $d$ between the transmitter and the receiver, $A_{\rm w_0}$ is the regression constant, and $\alpha$ is the slope of the linear regression, which depends on WDR and pressure head, $X_{\rm A}$ is a random variable representing the variations as noise in $A_{\rm w}(v)$ given as $X_{\rm A}\sim\mathcal{N}(0,\sigma_{\rm A}^2)$.
\begin{figure*}
 \center
 \vspace{-0.2 cm}
  \includegraphics[width=\textwidth]{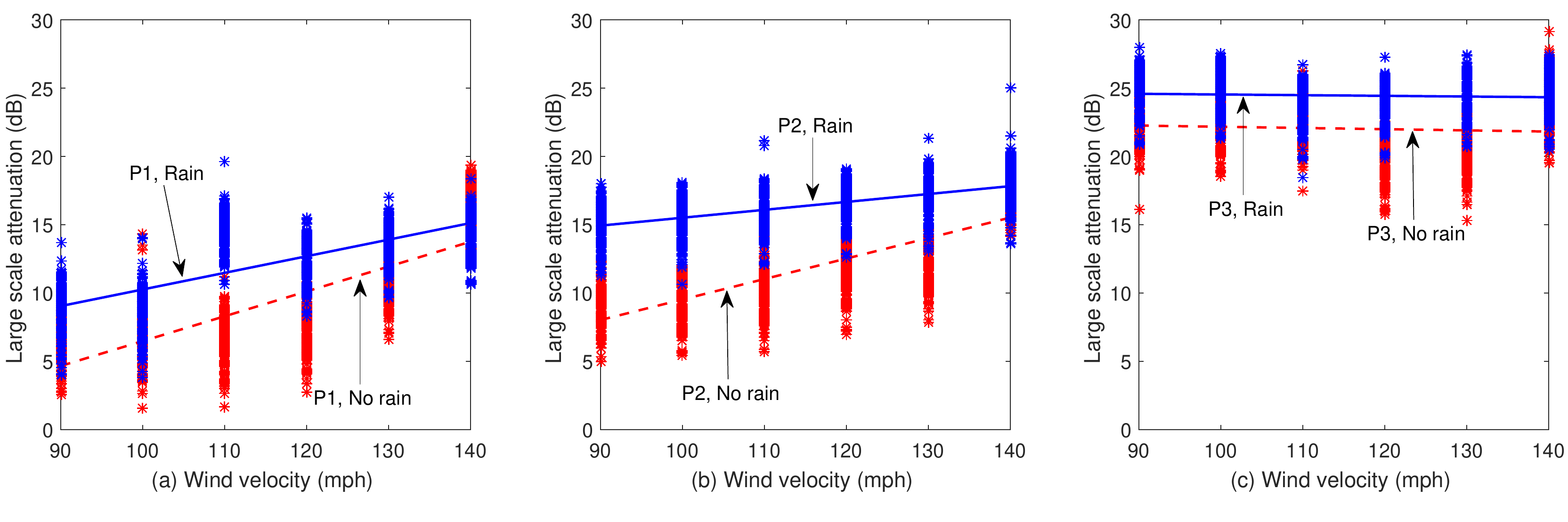}
    \caption{Large scale attenuation at different wind velocities for radio positions (a) P1 (LOS), (b) P2 (LOS), (c) P3 (NLOS). }\label{Fig:atten_los_nlos}
  \vspace{-0.4cm}
\end{figure*}
The large scale fading parameters that fits in to the model developed in this section using the empirical data explained in Section IV are given in Table~I. 
%\vspace{-0.3 cm}
\begin{table}[!t]
	\begin{center}
	\caption{Large scale fading parameters.}
    \vspace{-0.4 cm}
		\label{Table_I}
		\begin{tabular}{|P{1cm}|P{0.82cm}|P{0.82cm}|P{0.82cm}|P{0.82cm}|P{0.82cm}|P{0.82cm}|}
			\hline
			\textbf{Param.}&\textbf{P1, S1}&\textbf{P1, S2}&\textbf{P2, S1}&\textbf{P2, S2}&\textbf{P3, S1}&\textbf{P3, S2}\\
			\hline
			$\alpha$&$0.182$&$0.122$&$0.15$&$0.06$&$-0.01$&$-0.005$ \\
			\hline
		$A_{\rm w0}({\rm dB})$&$-11.7$&$-1.9$&$-5.5$&$9.73$&$23$&$25$ \\
			\hline
            $\sigma_{\rm A}({\rm dB})$&$12.39$&$10.09$&$13$&$11.53$&$18.32$&$16.75$\\
			\hline
			\end{tabular}
	\end{center}
\vspace{-0.2 cm}
\end{table}

Small scale fading for UWB signals in hurricane is due time dispersion of MPCs. There is no Doppler spread present due to static position of the transmitter and the receivers. The small scale fading amplitudes are better fitted with Nakagami distribution given as 
\vspace{-0.09 cm}
\begin{align}
F(y;m,\Omega) = \frac{2m^my^{2m-1}}{\Gamma(m) \Omega^m}\exp\Big(\frac{-my^2}{\Omega}\Big)~,
\end{align}
where $m$ is the shape factor for Nakagami distribution, $\Omega$ represents the spread controlling factor, and $\Gamma(m)$ represents the Gamma function. If $W$ represents the random variable $W\sim~{\rm Nakagami}\big(m,\Omega\big)$, then, we have~\cite{kolar}, 
$m =\frac{ E^2[W^2]}{{\rm Var}[W^2]}$, $\Omega = E[W^2]$, where the Nakagami-$m$ factor is log-normally distributed. Mean and variance of Nakagami-$m$ factor are represented by $\mu_{\rm mf}$ and $\sigma_{\rm mf}$, respectively, while the mean and variance of spread controlling factor $\Omega$ are represented by $\mu_{\rm sc}$ and $\sigma_{\rm sc}$ respectively. The small scale fading parameters obtained from empirical results that fit the model described above are provided in Table~II. 
%\vspace{-0.4 cm}
%\vspace{-0.3 cm}
\begin{table}[!t]
	\begin{center}		
		\caption{Small scale fading parameters.}\label{Table_III}
        \vspace{-0.5 cm}
		\begin{tabular}{|P{1cm}|P{0.82cm}|P{0.82cm}|P{0.82cm}|P{0.82cm}|P{0.82cm}|P{0.82cm}|}
			\hline
			\textbf{Param.}&\textbf{P1, S1}&\textbf{P1, S2}&\textbf{P2, S1}&\textbf{P2, S2}&\textbf{P3, S1}&\textbf{P3, S2}\\
			\hline
			$\mu_{\rm mf}(\rm dB)$&$-2.69$&$-2.96$&$-2.47$&$-3.25$&$-2.55$&$-2.92$ \\
			\hline	
			$\sigma_{\rm mf}(\rm dB)$&$-25.2$&$-25.7$&$-27.2$&$-31.5$&$-20.5$&$-19.9$ \\
			\hline	
			$\mu_{\rm sc}(\rm dB)$&$0.92$&$2.28$&$1.29$&$-3.07$&$-9.21$&$-10.6$ \\
			\hline	
			$\sigma_{\rm sc}(\rm dB) $&$1.21$&$6.26$&$-1.75$&$-3.85$&$-16.5$&$-27$ \\
			\hline	
		  \end{tabular}
		\end{center}
        \vspace{-0.4 cm}
 \end{table}

In case of LOS, the Nakagami-$m$ distribution can be approximated with a Rician distribution \cite{rician} given as 
\begin{align}
m &\cong \frac{(K+1)^2}{2K + 1}, \ \ G(y)  = \frac{y}{A_{0}}e^{\frac{y^2+x^2}{2A_{0}}I_{0}\frac{yx}{A_{0}}},
\end{align}
 where $K  = 10\log_{10}\big(\frac{x^2}{2A_{0}}\big)$, is the Rician \textit{K}-factor representing the ratio between power of LOS component to the scattered components, $G(y)$ represents the Rician distribution with $A_{0}$ representing the power in the diffuse MPCs, $x^2$ is the power in the dominant LOS component, and $I_{0}$ is the $0^{\rm th}$ order Bessel function. The power in the diffuse component $A_{0}$ in the hurricane conditions can be represented as
\begin{align}
A_{0}= A_{\rm b}(c_{\rm r0}b_{\rm r0} + c_{\rm p0}b_{\rm p0} + c_{\rm w0}v_{\rm w0}) + X_{\rm A_0}, \label{Eq:K_factor}
\end{align}
where $A_{\rm b}$ is the diffuse power of the base case, $c_{\rm r0}\geq0$, $c_{\rm w0}\geq0$ are constants proportional to WDR, and wind velocities respectively, $c_{\rm p0}\geq0$, is a constant that has higher value for lower pressure head regions and vice versa, $b_{\rm r0}$, $b_{\rm p0}$, and $v_{\rm w0}$ are the coefficient values for respective constants, where $b_{\rm r0}$ = [0,1] and $b_{\rm p0}$ = [0,1], representing either absence or presence of rain and lower pressure head respectively, and $v_{\rm w0}$ = [90:10:140] in our case, and $X_{\rm A_0}\sim\mathcal{N}(0,\sigma_{\rm A0}^2)$ is a Gaussian distributed random variable. We use linear least square error to calculate the values of the constants and distribution of $X_{\rm A_0}$ in (\ref{Eq:K_factor}) as follows: 
\begin{align}
[c_{\rm r0}\ c_{\rm p0}\ c_{\rm w0}]^{\rm T} = \sum_{ii=1}^{24}\sum_{jj=1}^{3}(M_{ii, jj}^{\rm T}M_{ ii,jj})^{-1}M_{ii,jj}^{\rm T}(A_{\rm 0}-A_{\rm b}),
\end{align}
 where $M$ is a matrix that contains the values [$b_{\rm r0} \ b_{\rm p0} \ v_{\rm w0}$]. 
 
The variation of power for diffuse and direct components in different scenarios averaged over scans is represented as $X_{\rm Df}\sim\mathcal{N}(\mu_{\rm Df},\sigma_{\rm Df}^2)$ and $X_{\rm Dr}\sim\mathcal{N}(\mu_{\rm Dr},\sigma_{\rm Dr}^2)$ respectively. The variations in \textit{K}-factor is represented  by $X_{\rm K}\sim\mathcal{N}(\mu_{\rm K},\sigma_{\rm K}^2)$. In case of NLOS, $K=0$ and the Rician distribution converges to Rayleigh distribution. 

%\vspace{0.1 cm}
%\textbf{ Remark 5:} The dominant LOS component power is less effected compared to diffuse component during hurricane. The values of $c_{\rm w_0}$ and $c_{\rm p_0}$ in (\ref{Eq:K_factor}) are very small for smaller wind velocity variations and become dominant after wind velocities approaching 140 mph. 

%\nocite{key1,key2,key3,key4}

%\begin{figure}[!h]
 %\centering
 % \includegraphics[width=\columnwidth]{PDP.pdf}
  %\caption{Normalized PDP at different wind velocities for (a) P1 (No rain) , (b) P2 (No rain), (c) P3 (No rain), (d) P1 (Rain), (e) P2 (Rain), (f) P3 (Rain).}\label{Fig:PDP_los}
%\end{figure}

\section{Empirical Results}
This section provides an analysis of large and small scale fading, and the statistics of the MPCs, based on the experiments described in Section II. After collecting the data, post processing of the measurement data is carried out in Matlab. 
\subsection{Large and Small Scale Fading Analysis}
Large scale attenuation in our case is a measure of reduction in the received energy during hurricane conditions. Large scale fading for different radio positions and respective scenarios as a function of wind velocity, based on (\ref{Eq:Largescale_1}) and (\ref{Eq:Largescale}), is shown in Fig.~\ref{Fig:atten_los_nlos}. Results show that in general, higher attenuation is observed at higher wind speeds for LOS scenarios, while for NLOS scenarios no critical impact of wind speed is observed.  Comparison of our results with related literature \cite{Weather_new3,uwb4,uwb6} indicates that we observe higher attenuation in our experiment, which may be due to following reasons.

First, change in the ambient conditions under different wind velocities can lead to attenuation without rain. This is due to ambient noise that varies with wind flows and pressure heads. At P2, we have lower pressure head that introduces higher wind gusts. Also, at P2, we observe bouncing and scattering of strong wind currents from the obstacle. This results in additional ambient noise at P2 due to exposure to more agitated molecules resulting in more attenuation. Second, the mechanical turbulences of communication, tie down and relaying equipment introduces variations in the received power. The highest mechanical turbulences are experienced by the exposed antennas. With transmitter and receivers at same height, we observe additional attenuation due to non-alignment of the bore-sight of antennas \cite{TD}. Another reason is the limitation of the communication equipment itself. For a given receiver sensitivity, a reduction in the SNR due to ambient noise and mechanical turbulences during hurricane conditions causes the packets to be dropped at the receiver. This implies a more lossy channel. Finally, in case of WDR, the wind intensity we considered is much higher than considered in the literature \cite{uwb4,Weather_new3}. This high intensity rain driven by high wind velocities form a kind of water wall between the transmitter and the receiver. Due to higher dielectric constant and loss tangent of the water as compared to rain, we observe additional attenuation during rain. This attenuation is accompanied by scattering, refraction and diffraction of EM waves and higher ambient noise from water droplets that depends on the droplet size distribution. Additionally, the accumulation of water on different surrounding objects result in reduced reflections of incident EM energy due to absorption.

In case of NLOS measurements in Fig.~\ref{Fig:atten_los_nlos}(c), we have higher attenuation as the energy is compared with the standard LOS path scenario. In addition to the effects explained above, the absence of dominant LOS component results in more attenuation as compared to LOS.  
\begin{figure}[!h]
	\centering
	\includegraphics[width=\columnwidth]{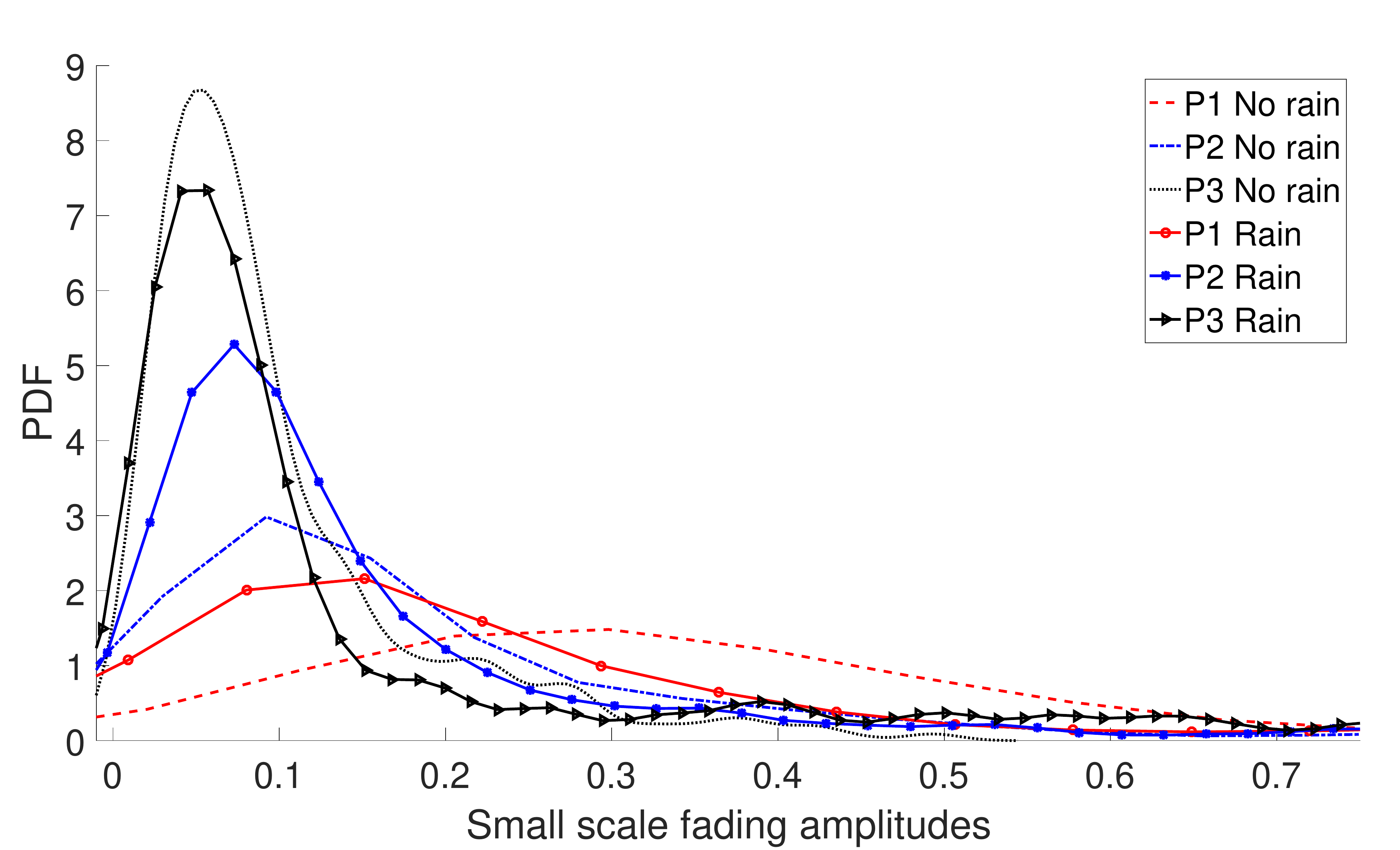}
	\caption{Probability density function of small scale fading amplitudes for wind velocity $v$=90 mph for different radio positions and respective scenarios.}\label{Fig:small_scale_fading}
     \vspace{-0.1cm}
\end{figure}

The probability density function (PDF) of small scale fading amplitudes at different radio positions and respective scenarios is shown in Fig.~\ref{Fig:small_scale_fading} for wind velocity of 90 m/s. A general trend is that with WDR, there is decrease of mean and variance of the PDF for both LOS and NLOS paths. This reduction is higher for P2 as compared to P1 for the LOS path. We can deduce that with rain and low pressure head, we expect more attenuation and lower small scale amplitude fluctuations. The mean and variance of the LOS path is greater than the NLOS path. This is validated by the Nakagami-$m$ factor and spread controlling factor $\Omega$ values given in Table~II. It can be observed that for NLOS, the distribution fits closely to Raleigh. Similar results are obtained for other wind velocities. 

\begin{figure}[!h]
	\centering
	%\vspace{-.7cm}
	\includegraphics[width=\columnwidth]{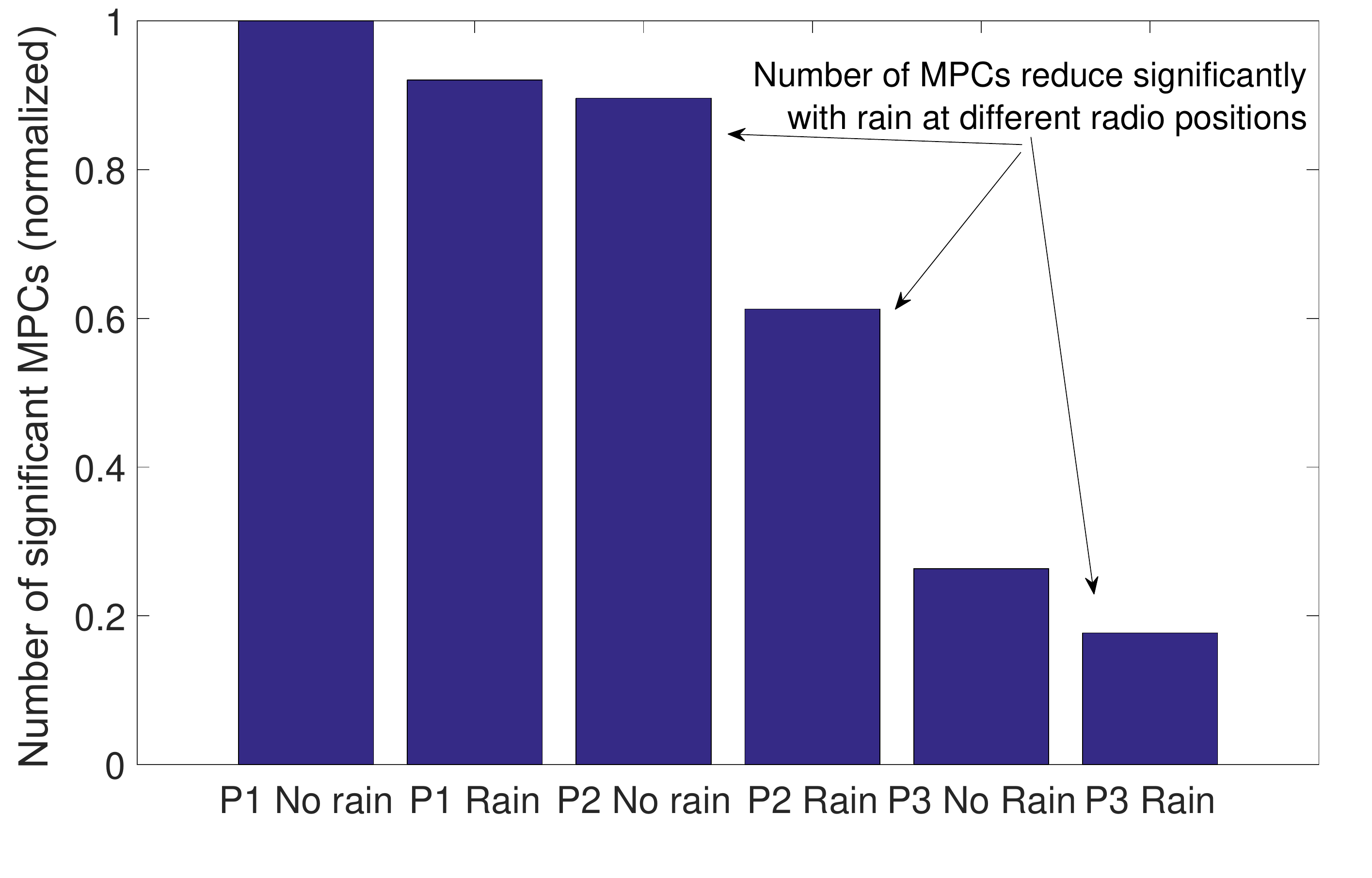}
	%\vspace{-0.8cm}
	\caption{Number of significant MPCs above threshold averaged over scans for different radio positions and corresponding scenarios (normalized to the number of significant MPCs for no rain scenario at P1).
     \vspace{-0.25cm}}\label{Fig:Significant_paths}
\end{figure}
\subsection{Multipath Channel Analysis}
The MPCs from empirical data are analyzed to determine the behavior of the channel and validate the stochastic model developed in Section IV. Significant MPCs are characterized to have amplitude greater than the threshold set at 15\% of the maximum amplitude for a given scenario as shown in Fig.~\ref{Fig:Significant_paths}. It can be observed that number of significant MPCs are affected by radio position and rain in case of LOS due to the effects discussed earlier resulting in weakening of MPC amplitudes especially during rain. In case of NLOS, we observe higher reduction as the arriving MPCs are weak. The multipath statistical channel parameters obtained from empirical data, that fit the model for MPCs propagation through hurricanes, developed in Section III are given in Table~III.

\begin{figure}[!t]
	\centering
	\includegraphics[width=\columnwidth]{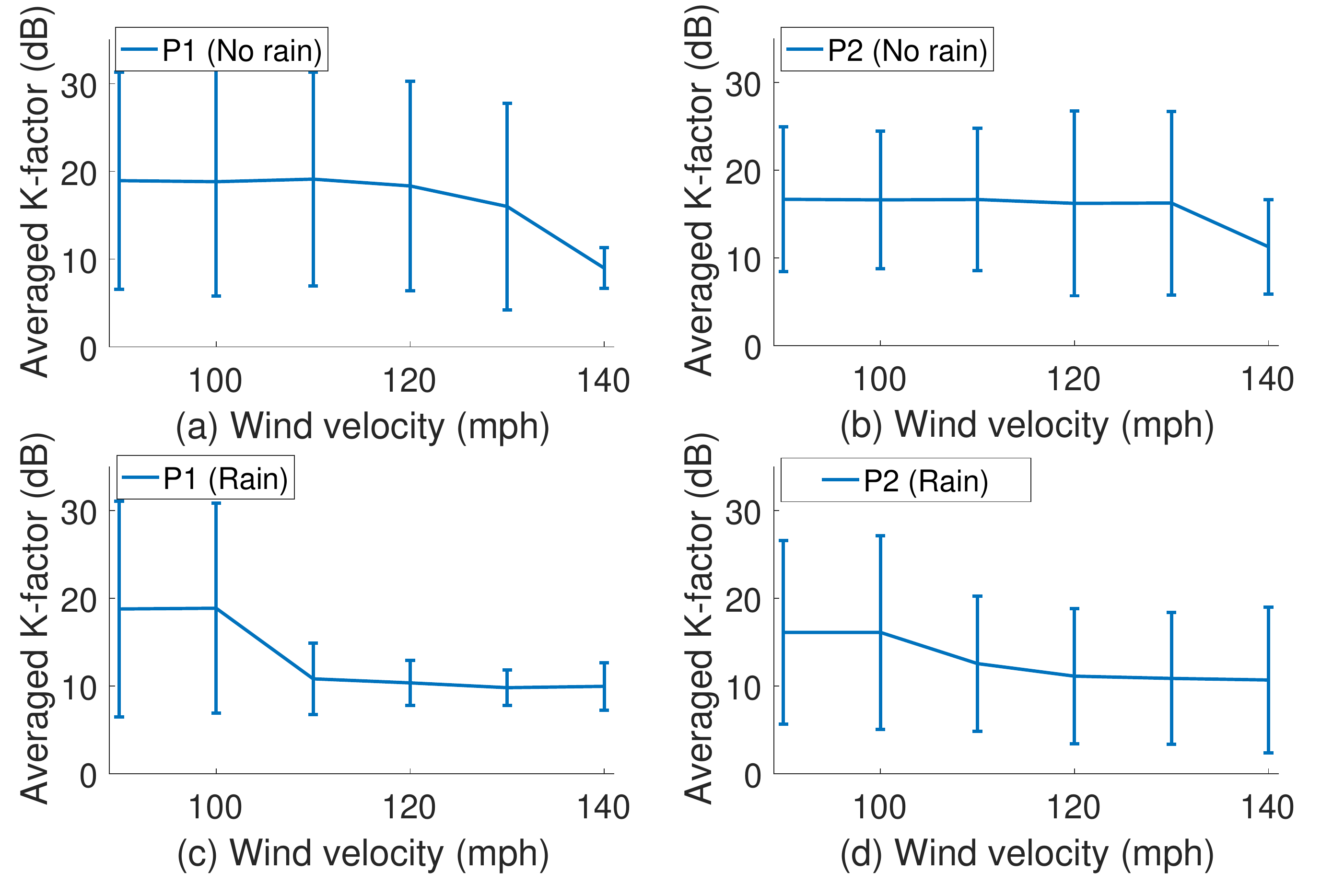}
	\caption{ $K$-factors for different wind velocities at different radio positions (a) P1 (no rain), (b) P2 (no rain), (c) P1 (rain), (d) P2 (rain).
     \vspace{-0.4cm}}\label{Fig:K_factor}
\end{figure}

The Rician $K$-factor plot for the LOS path averaged over multiple scans is shown in Fig.~\ref{Fig:K_factor}. From~(\ref{Eq:K_factor}), a decrease in \textit{K}-factor is observed with the increase in the wind velocities similar to \cite{uwb6} and \cite{Weather_new4}, especially in case of WDR. It can be observed that there is less effect of pressure head on the $K$-factor without rain, as the dominant LOS component is less affected by wind velocities as shown in Fig.~\ref{Fig:K_factor}(a) and Fig.~\ref{Fig:K_factor}(b). The $K$-factor remains almost constant and changes significantly only at 140 mph indicating that ratio of dominant LOS component power and diffuse components power remains proportional except at 140 mph where the dominant LOS component power has reduced significantly. In case of WDR shown in Fig.~\ref{Fig:K_factor}(c) and Fig.~\ref{Fig:K_factor}(d), we observe a significant reduction in the $K$-factor at 110 mph indicating that due to rain the dominant LOS component is reduced earlier as compared to no rain. For the rest of the wind velocities, the $K$-factor remains same indicating that there is proportional reduction in the dominant LOS component and diffuse component. 

\begin{table}[!t]
	\begin{center}
		\caption{Channel model parameters.}\label{Table_IV}
        \vspace{-0.4 cm}
		\begin{tabular}{@{} |P{1cm}|P{0.82cm}|P{0.82cm}|P{0.82cm}|P{0.82cm}|P{0.82cm}|P{0.82cm}| @{}}
			\hline
			\textbf{Param.}&\textbf{P1, S1}&\textbf{P1, S2}&\textbf{P2, S1}&\textbf{P2, S2}&\textbf{P3, S1}&\textbf{P3, S2}\\			
			\hline
			$\overline N$&$5.2$&$4$&$3.33$&$3.1$&$1.67$&$1.67$ \\
			\hline
			$\gamma({\rm 1/ns})$&$0.11$&$0.06$&$0.043$&$0.027$&$0.017$&$0.017$\\
			\hline
			$\zeta({\rm 1/ns})$&$16.32$&$12.6$&$9.32$&$5.61$&$2.33$&$0.3$\\
			\hline
			$\Lambda({\rm ns})$&$2.3$&$2.45$&$2.38$&$2.47$&$1.72$&$1.86$ \\
			\hline
			$\lambda({\rm ns})$&$0.8$&$0.91$&$0.82$&$0.85$&$0.54$&$0.61$ \\
			\hline
            $\sigma_{\rm a}(\rm dB)$&$23.14$&$21.22$&$21.56$&$18.51$&$15.12$&$13.2$ \\
			\hline
			$\sigma_{\rm \overline N}$&$1.59$&$.98$&$.836$&$.837$&$.53$&$.516$\\
			\hline	
			$\sigma_{\rm c}({\rm ns})$&$16.4$&$34.5$&$43.93$&$49.47$&$51.63$&$52.14$\\
			\hline
			$\sigma_{\rm m}(ns)$&$0.287$&$0.268$&$0.277$&$0.265$&$0.28$&$0.253$\\
			\hline
			$\sigma_{\rm P}({\rm dB})$&$21.3$&$20.1$&$20.03$&$19.72$&$17.41$&$14.1$ \\
			\hline
			$\sigma_{\rm mp}({\rm dB})$&$55.14$&$38.44$&$33.22$&$26.77$&$20.17$&$15.22$\\
			\hline
			$\sigma_{\rm R}({\rm dB})$&$-$&$5.77$&$-$&$6.39$&$-$&$7.31$\\
			\hline			
			$\sigma_{\rm A0}({\rm dB})$&$28.97$&$10.45$&$8.44$&$.219$&$6.46$&$.959$\\
			\hline
			$\sigma_{\rm Df}({\rm dB})$&$18.83$&$17.79$&$17.6$&$15.3$&$12.78$&$11.45$\\
			\hline
            $\mu_{\rm Df}({\rm dB})$&$16.19$&$15.17$&$15.28$&$13.54$&$11.78$&$10.9$\\
			\hline         
            $\sigma_{\rm Dr}({\rm dB})$&$11.53$&$10.97$&$10.64$&$9.6$&$-$&$-$\\
			\hline
            $\mu_{\rm Dr}({\rm dB})$&$33.3$&$31$&$30.7$&$27.86$&$-$&$-$\\
			\hline
            $\sigma_{\rm K}({\rm dB})$&$14.64$&$14.5$&$11.2$&$6.44$&$-$&$-$\\
			\hline
			$\mu_{\rm K}({\rm dB})$&$17.69$&$14.43$&$16$&$13.79$&$-$&$-$\\
			\hline 
	\end{tabular}
		\end{center}
\vspace{-0.4 cm}
\end{table}

\section{Conclusions and Future Work}
In this work, we have conducted UWB propagation channel measurements in different hurricane conditions. The effects of wind velocity and WDR at different pressure head regions are monitored in LOS and NLOS communication paths. Based on the measurements, a UWB channel model for hurricanes is developed. It is observed that UWB communications under different wind velocities, pressure heads and WDR for LOS and NLOS paths in a hurricane introduce different effects on the channel model parameters (large scale, small scale and MPCs propagation). The proposed model can help in improved design of communications systems in hurricane conditions. Our future work includes developing channel models in hurricane conditions for millimeter wave communications where more severe attenuation is expected to occur.

\bibliographystyle{IEEEtran}
%\bibliography{VTC_2017}

% Generated by IEEEtran.bst, version: 1.14 (2015/08/26)

\end{document}